\documentclass[iop]{emulateapj-rtx4}

\usepackage{mathrsfs }
\usepackage{natbib}
\usepackage{amssymb}
\usepackage{ulem} 
\newcommand{\secpoint}{\mbox{$''\mskip-7.6mu.\,$}}

\begin{document}

\title{Q1549-C25: A Clean Source of Lyman-Continuum Emission at $z=3.15$\altaffilmark{1}}

\author{
 Alice E. Shapley,\altaffilmark{2}
 Charles C. Steidel,\altaffilmark{3}
 Allison L. Strom,\altaffilmark{3}
 Milan Bogosavljevi\'c,\altaffilmark{4}
 Naveen A. Reddy,\altaffilmark{5}
 Brian Siana,\altaffilmark{5}
 Robin E. Mostardi,\altaffilmark{6}	
 Gwen C. Rudie\altaffilmark{7}	
 }

\altaffiltext{1}{Based on data obtained at the W.M. Keck Observatory, which is operated as a scientific partnership among the California Institute of Technology, the University of California,  and the National Aeronautics and Space Administration, and was made possible by the generous financial support of the W.M. Keck Foundation.}
\altaffiltext{2}{Department of Physics and Astronomy, University of California, Los Angeles, 430 Portola Plaza, Los Angeles, CA 90095, USA}
\altaffiltext{3}{Cahill Center for Astronomy and Astrophysics, California Institute of Technology, 1216 East California Boulevard., MS 249-17, Pasadena, CA 91125, USA}
\altaffiltext{4}{Astronomical Observatory, Volgina 7, 11060 Belgrade, Serbia}
\altaffiltext{5}{Department of Physics and Astronomy, University of California, Riverside, 900 University Avenue, Riverside, CA 92521, USA}
\altaffiltext{6}{Physics Department, Los Positas College, 3000 Campus Hill Drive Livermore CA 94551, USA}
\altaffiltext{7}{Carnegie Observatories, 813 Santa Barbara Street, Pasadena, CA 91101, USA}
\email{aes@astro.ucla.edu}

\shortauthors{Shapley et al.}


\shorttitle{A Clean Source of LyC Emission}


\begin{abstract} 
We present observations of Q1549-C25, an $\sim L^*$ star-forming galaxy at $z=3.15$
for which Lyman-continuum (LyC) radiation is significantly detected in deep
Keck/LRIS spectroscopy. We find no evidence for contamination from a lower-redshift
interloper close to the line of sight in the high signal-to-noise
spectrum of Q1549-C25. Furthermore, the morphology of Q1549-C25
in $V_{606}$, $J_{125}$, and $H_{160}$ {\it Hubble Space Telescope} ({\it HST}) imaging
reveals that the object consists of a single, isolated component within $1"$.
In combination, these data indicate Q1549-C25 as a {\it clean} spectroscopic detection of LyC
radiation, only the second such object discovered to date at $z\sim 3$. We model the 
spectral energy distribution (SED) of Q1549-C25, finding evidence for negligible
dust extinction, an age (assuming continuous star formation) of $\sim 1$~Gyr,
and a stellar mass of $M_*=7.9\times 10^{9} M_{\odot}$. Although it is not
possible to derive strong constraints on the absolute escape fraction of LyC emission,
$f_{\mbox{esc}}(\mbox{LyC})$, from a single object, we use simulations
of intergalactic and circumgalactic absorption to infer $f_{\mbox{esc}}(\mbox{LyC}) \geq 0.51$
at 95\% confidence. The combination of deep Keck/LRIS spectroscopy and {\it HST}
imaging is required to assemble a larger sample of objects like Q1549-C25,
and obtain robust constraints on the average $f_{\mbox{esc}}(\mbox{LyC})$ at $z\sim 3$
and beyond.

\end{abstract} 



\section{Introduction}
\label{sec:introduction}

The  escape fraction of Lyman-continuum
(LyC) photons from galaxies is a crucial
component of models of the reionization of the universe.
In recent reionization models, many reasonable assumptions have
been adopted or constraints derived for the LyC escape fraction
($f_{\mbox{esc}}(\mbox{LyC})$)
\citep[e.g.,][]{robertson2015,finkelstein2012}. However,
such constraints are typically indirect \citep{kuhlen2012}
and do not substitute for the actual detection of ionizing photons leaking
from galaxies. Because of the increasing intergalactic medium (IGM)
optical depth at higher redshifts, it is not possible
to directly measure escaping ionizing radiation from
galaxies much beyond $z\sim 3$, let alone during the
epoch of reionization \citep{vanzella2012}.
Therefore, such measurements must be performed at
$z\leq 3.5$, the highest redshift at which IGM
absorption does not destroy the signal of interest.

Direct measurements of LyC emission can be used for estimating
the average $f_{\mbox{esc}}(\mbox{LyC})$ at $z\sim 3$,
and the relationships between $f_{\mbox{esc}}(\mbox{LyC})$ and other galaxy 
properties. Determining these relationships is crucial for translating
measurements of non-ionizing radiation 
from galaxies during the epoch of reionization
into an estimate of their contribution to the ionizing budget.

Robust detections of LyC emission have recently been obtained
for several low-redshift galaxies using COS
on the {\it Hubble Space Telescope} ({\it HST}) \citep[e.g.,][]{borthakur2014,izotov2016,leitherer2016}.
At $z\sim 3$, both spectroscopic \citep[e.g.,][]{shapley2006} and ground-based
and {\it HST} imaging techniques have been used to measure LyC emission
\citep[e.g.,][]{nestor2011,mostardi2013,grazian2016,vanzella2012}.
In ground-based $z\sim 3$ LyC observations, contamination by lower-redshift interlopers
is a serious concern. When contaminated, the flux at $\sim 3500$\AA\ observed
does not consist of LyC at $z\sim 3$, but rather
non-ionizing UV-continuum from a lower-redshift
source near the line of sight. High-spatial-resolution observations 
(e.g., with {\it HST}) are required to rule out contamination \citep{vanzella2012}. To date, there is only one object
with a spectroscopic detection of LyC at $z\sim 3$ and uncontaminated {\it HST} morphology, i.e.,
{\it Ion2} at $z=3.21$ \citep{debarros2016,vanzella2016}.

\begin{figure*}[ht!]
\centering
\includegraphics[width=0.95\textwidth]{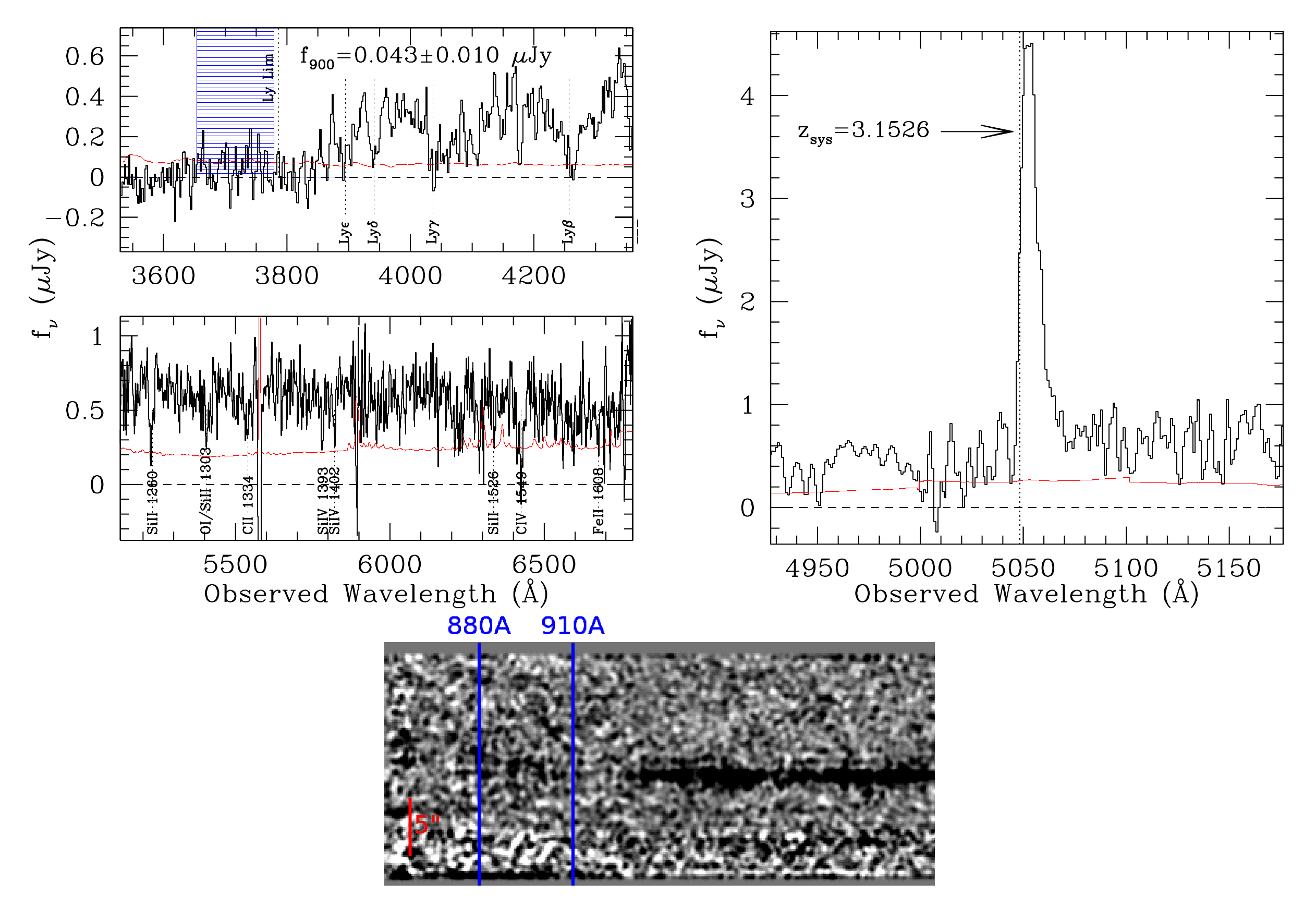}
\caption{{\bf Top:} Deep LRIS spectrum of Q1549-C25. The spectrum is presented in the observed
frame, and the flux-density units are $\mu$Jy. The error spectrum is plotted
in red. Blue- and red-side spectra are joined
together at the 5000~\AA\ dichroic cut-off. {\bf Top left:} The top panel shows the LyC range ($880-910$\AA),
indicated as a blue shaded region, and Lyman-series absorption lines. $f_{900}$ is evaluated
as the mean flux-density in the LyC region. The bottom panel shows the spectral
range longwards of Ly$\alpha$, containing several labeled interstellar metal absorption lines.
{\bf Top right:} Zoomed-in spectrum of Ly$\alpha$ emission. The Ly$\alpha$ profile
profile is characterized by a single emission peak offset relative to the
systemic redshift of Q1549-C25, indicated by the vertical dotted line.
{\bf Bottom:} Two-dimensional spectrum of Q1549-C25, with the LyC region bracketed by vertical
blue lines at rest wavelengths of 880 and 990\AA. Wavelength increases
from left to right and the spatial scale is indicated with a vertical bar of 5" in extent.
The spectrum has been lightly smoothed by a Gaussian kernel with 3-pixel radius.
}
\label{fig:c25spec}
\end{figure*}

\begin{figure}
\centering
\includegraphics[width=0.5\textwidth]{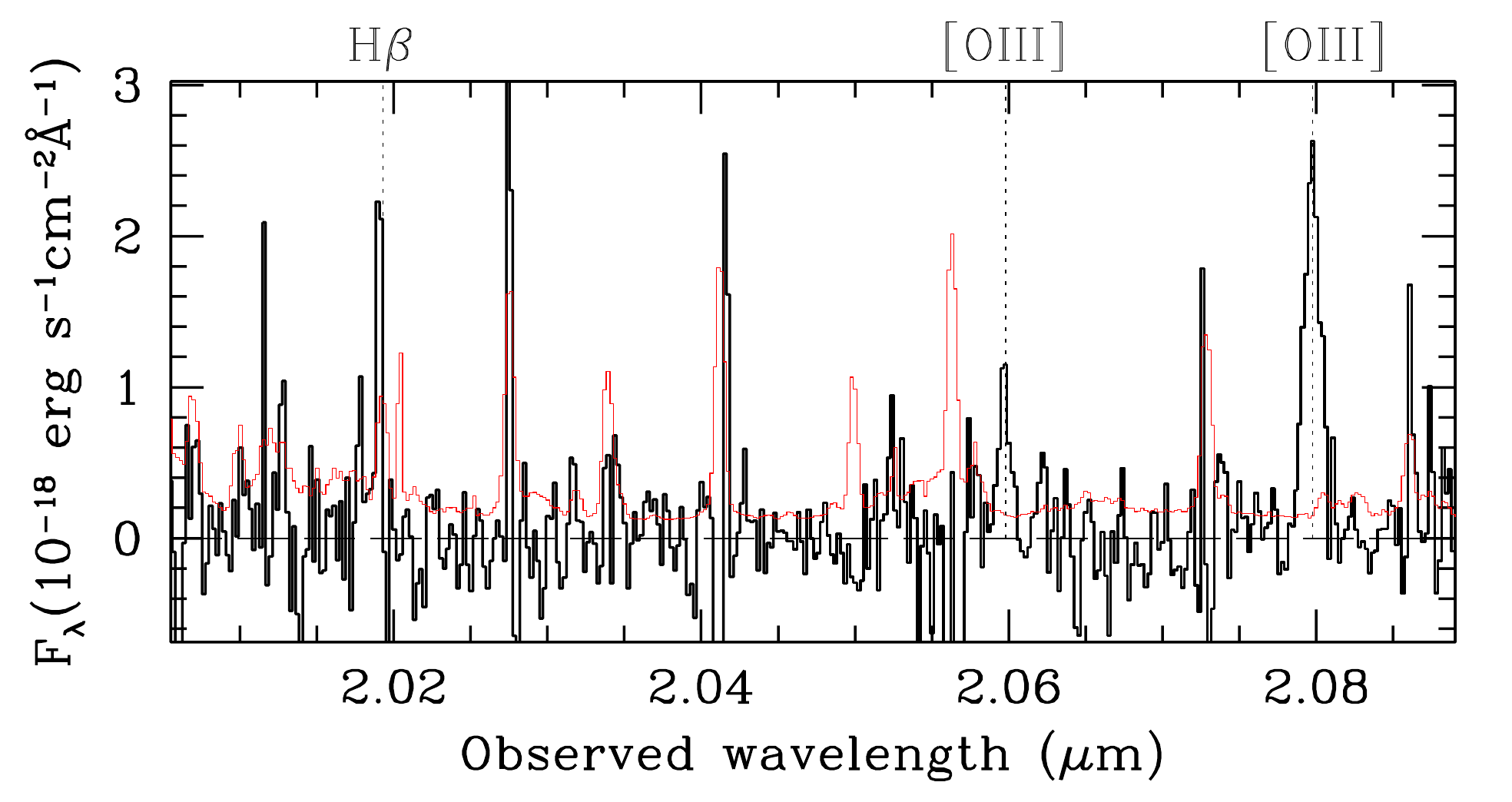}
\caption{MOSFIRE $K$-band spectrum of Q1549-C25. The spectrum is presented
in the observed frame and the flux-density units are 
$10^{-18}\mbox{ ergs s}^{-1}\mbox{ cm}^{-2}\mbox{ \AA}^{-1}$. The error
spectrum is plotted in red. The [OIII]$\lambda5007$
and [OIII]$\lambda 4959$ lines are clearly detected, while H$\beta$ falls
on a sky line. The combined [OIII]+H$\beta$ emission-line flux is 
$\leq 5.2\times10^{-17} \mbox{ergs s}^{-1}\mbox{ cm}^{-2}$, corresponding to a rest-frame
equivalent width of $W_{(\mbox{[OIII]+H}\beta),0}\leq256$\AA. 
The systemic redshift based on [OIII] emission is $z_{sys}=3.1526$.
}
\label{fig:c25spec-mosfire}
\end{figure}

As we describe here, we have obtained deep Keck/LRIS spectroscopy for a large
sample of Lyman Break Galaxies (LBGs) at $z\sim 3$, including coverage
of the LyC region (Steidel et al., in prep.). Roughly 10\% of these galaxies show spectroscopic
detections of LyC radiation. One of them, Q1549-C25, is also covered
by multi-wavelength {\it HST} imaging, from which we determine that
the galaxy is unaffected by low-redshift contamination and represents a clean
detection of LyC emission.
In \S\ref{sec:observations}, we describe our spectroscopic and imaging
observations.  In \S\ref{sec:results}, we present the spectroscopic
detection of LyC emission in Q1549-C25, along with the galaxy's multi-wavelength morphology and
stellar population parameters.
Finally, in \S\ref{sec:discussion}, we discuss the implications for estimating
$f_{\mbox{esc}}(\mbox{LyC})$, compare the properties of $z\sim 3$ galaxies detected in LyC,
and consider the outlook for LyC observations at high redshift.
Throughout, we adopt cosmological parameters of
$H_0=70 \mbox{ km  s}^{-1} \mbox{ Mpc}^{-1}$, $\Omega_M = 0.30$, and
$\Omega_{\Lambda}=0.7$.

\begin{figure*}
\centering
\includegraphics[width=0.95\textwidth]{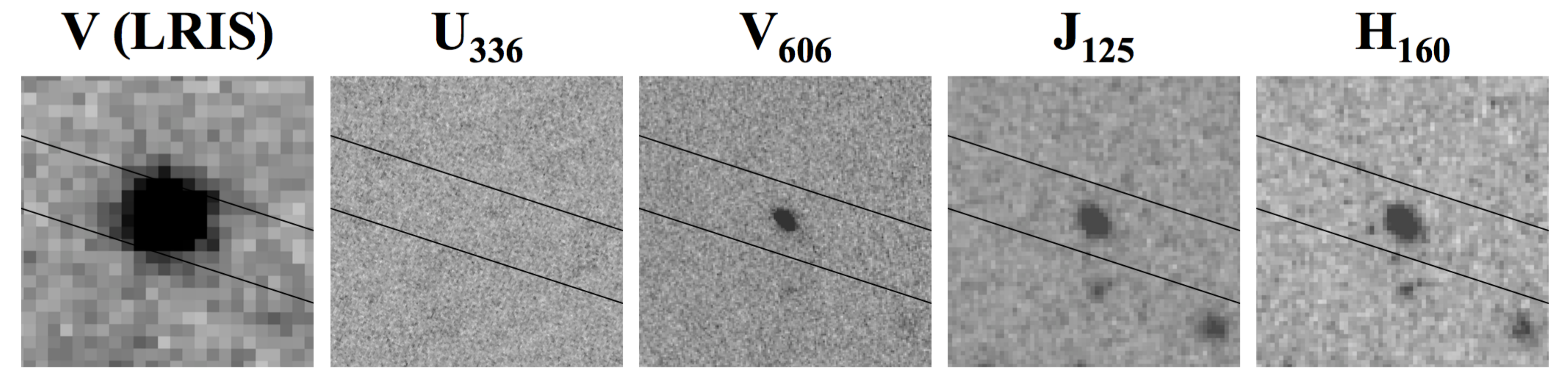}
\caption{$5"\times5"$ postage stamp images of Q1549-C25. From left
to right, we show ground-based $V$ band, $U_{336}$, $V_{606}$, $J_{125}$, and $H_{160}$,
oriented with north up and east to the left. The LRIS slit is overlaid on each postage
stamp.  These images indicate that C25 consists of a single component within a $1"$
radius. The faint galaxy 1\secpoint3 to the south does not contribute to
the flux measured in the LRIS spectrum. The lack of $U_{336}$ detection for Q1549-C25 is
consistent with the spectroscopic detection of $f_{900}$, given the depth of the $U_{336}$ imaging.
}
\label{fig:c25-panelfig}
\end{figure*}

\section{Observations \& Methods}
\label{sec:observations}

\subsection{Keck Spectroscopy}
\label{sec:observations-keck}
As described in Steidel et al. (in prep.),
we used the LRIS spectrograph on the Keck~I telescope to assemble
a sample of 136 galaxies at $2.72\leq z\leq 3.52$
($\langle z\rangle = 3.05\pm0.18$) with deep rest-UV spectra covering the LyC
region (rest-frame $880-910$\AA). The galaxy Q1549-C25 has coordinates of
R.A.=15:52:06.07 and decl.=+19:11:28.4 (J2000), ${\cal R}_{AB}=24.83$ 
\citep[i.e., roughly $L^*$;][]{reddy2008}, and falls in the HS1549+1919 field,
one of the survey fields of the Keck Baryonic Structure Survey \citep[KBSS;][]{rudie2012,steidel2014}.
This object was originally identified as a $z\sim 3$ photometric candidate and confirmed
at $z_{Ly\alpha}=3.16$ using, respectively, LRIS $U_nG{\cal R}$ imaging and spectroscopy
\citep{steidel2003,reddy2008}. In subsequent deep LyC spectroscopy, Q1549-C25 was observed through
a multi-object slitmask with 1\secpoint2 slits, at a sky position angle of $\theta=72^{\circ}$. 
Data were collected in April and June 2008, for a total exposure time of 8.4 hours under photometric
conditions and seeing $\sim$0\secpoint5$-$0\secpoint7. LRIS was configured with the ``d500" dichroic,
sending wavelengths shorter and longer than $5000$\AA, respectively, to the blue and red
channels, where they were dispersed, respectively, by a 400 lines/mm grism blazed at 3400\AA,
and 600 lines/mm grating blazed at 5000\AA. 

LRIS spectroscopic data were reduced as described in Steidel et al. (in prep.). 
In brief, individual two-dimensional spectra were flatfielded, cut out, rectified, corrected
for non-uniform slit illumination, background-subtracted, combined, extracted to one dimension,
and wavelength and flux-calibrated. The spectra were also dereddened for Galactic extinction,
using $E(B-V)=0.045$ for the HS1549+1919 field \citep{schlegel1998}. 
When trying to detect the faint signal of LyC emission, 
it is crucial to minimize and quantify the systematic and statistical uncertainties associated with 
these steps. We used careful tests for residual systematic errors in background
subtraction to establish that the zero flux level was robustly estimated (see Steidel et al., in prep
for details),\footnote{As described in Steidel et al. (in prep.), we performed multiple
careful tests of our background subtraction. In one such test, we
created a stack of 10 two-dimensional spectra with LyC detections
(including the spectrum of Q1549-C25), registering the individual
spectra at the position of each target galaxy. From this two-dimensional stack, we
extracted spectra over the wavelength regions $3450-3650$\AA\ and $4000-4100$\AA,
including all pixels along the slit except those within $\pm 4$" of the
stacked target position.  The resulting pixel flux distributions in each of these wavelength
regions are centered on zero counts, in agreement with pure Gaussian
noise, with a standard deviation as expected by our noise model
including photon counting statistics and detector read noise.}
a significant improvement over previous analyses \citep[e.g.,][]{shapley2006}.

A Keck/MOSFIRE $K$-band spectrum was obtained for Q1549-C25 in May 2016, for a total
of 1.5 hours in photometric conditions with 0\secpoint7 seeing. 
Observations were performed and data reduced as described in \citet{steidel2014}.

\subsection{{\it HST} and Other Imaging}
\label{sec:observations-hst}

We have obtained deep {\it HST} imaging
in two pointings in the HS1549+1919 field \citep{mostardi2015}. Each pointing
is covered by WFC3/UVIS $U_{336}$ (5 orbits), ACS/WFC $V_{606}$ (5 orbits), WFC3/IR $J_{125}$ (3 orbits)
and $H_{160}$ (3 orbits). These pointings were chosen to optimize the number of 
LBGs and Ly$\alpha$ emitters (LAEs) with apparent LyC emission detections
inferred from $3420$\AA\ narrow-band imaging \citep{mostardi2013}. Despite not being
detected at $3420$\AA, Q1549-C25 has coverage in
all four {\it HST} bands, enabling a careful analysis of its multi-wavelength morphology.

In addition, there is ground-based optical and near-IR imaging, as well as 
{\it Spitzer}/IRAC photometry for Q1549-C25. These include the original Keck/LRIS 
$U_nG{\cal R}$ plus $V$-band imaging, $J$ and $K_s$ from Palomar/WIRC
\citep{reddy2012}, $K_s$ and medium-band $J_1$, $J_2$, $J_3$, $H_{short}$, and $H_{long}$ from Magellan/FourStar, 
and IRAC channels 2 ($4.5\mu m$) and 4 ($8.0\mu m$). Q1549-C25 is detected in all
bands except $U_{336}$, $U_n$, $J$, $H_{long}$, and IRAC channel 4.

\section{Results}
\label{sec:results}

\subsection{The Direct Detection of LyC}
\label{sec:results-lyc}

Q1549-C25 is one of 13 galaxies in the LRIS LyC sample with
a $\geq3\sigma$ detection of $f_{900}$, the average flux density at rest-frame $880-910$\AA. The
rest-frame UV spectrum of Q1549-C25 is shown in Figure~\ref{fig:c25spec}. The top left
panel highlights the LyC region, in which we measure
$f_{900}=0.043\pm0.010$~$\mu$Jy, corresponding to an AB magnitude
of $m_{900}=27.33\pm0.26$. The bottom panel shows the two-dimensional spectrum
of Q1549-C25 over the LyC region, where a faint signal is apparent.
The average non-ionizing UV flux density is estimated
from the LRIS spectrum over the rest-frame range $1480-1520$\AA, yielding 
$f_{1500}=0.523\pm0.019$~$\mu$Jy. Combining these measurements, we find
$f_{900}/f_{1500}=0.08\pm0.02$ for the ratio of ionizing to non-ionizing flux
density.

The rest-frame UV spectrum of Q1549-C25 features strong  Ly$\alpha$
emission (rest-frame equivalent
width, $W_{Ly\alpha,0}=15$~\AA) at $z_{Ly\alpha}=3.156$ (see Figure~\ref{fig:c25spec}, top right)
and several low- 
(Si~II~$\lambda1260$, OI+Si~II~$\lambda1303$, C~II~$\lambda1334$,
Si~II~$\lambda1526$, Fe~II~$\lambda1608$)
and high-ionization (Si~IV~$\lambda\lambda1393,1402$, C~IV~$\lambda\lambda 1548,1550$)
interstellar metal absorption features at $z_{abs}=3.149$ (top left panel). The difference between
Ly$\alpha$ emission and interstellar absorption redshifts arises due to large-scale outflow motions
in the ISM of Q1549-C25 \citep[e.g.,][]{pettini2001,shapley2003}. 
The systemic redshift, $z_{sys}=3.1526$, is measured from the [OIII]$\lambda 5007$
emission centroid in the MOSFIRE spectrum (Figure~\ref{fig:c25spec-mosfire}).

Multiple Lyman-series absorption lines are also detected in the spectrum of Q1549-C25, although
their profiles may be contaminated by absorption from intervening Ly$\alpha$ forest
features. Given the high signal-to-noise of the spectrum, it is finally worth noting that
there is no evidence of a spectroscopic ``blend" with a lower-redshift
object along the line of sight. Such contamination would have appeared in the form
of absorption or emission features corresponding to an additional, lower redshift
\citep[see, e.g., Figure 5 of][]{siana2015}.

\subsection{Multi-wavelength Morphology}
\label{sec:results-morph}

Q1549-C25 is the only galaxy in our LRIS LyC sample with a significant
LyC detection for which multi-wavelength {\it HST} imaging also exists. 
Figure~\ref{fig:c25-panelfig} shows
ground-based $V$-band, along with {\it HST} $U_{336}V_{606}J_{125}H_{160}$
imaging. In contrast to the majority of $z\sim 3$ apparent sources of LyC
emission \citep{vanzella2012,mostardi2015,siana2015}, 
Q1549-C25 consists of a single source of emission at {\it HST} resolution, with no nearby sources of potential
contamination to the LRIS spectrum. The closest source to Q1549-25 is at a radial
separation of 1\secpoint3 in the southern direction, well outside the LRIS slit, 
significantly fainter than Q1549-C25 at all wavelengths, and undetected in $U_{336}$.
Although there are two apparent positive fluctuations at separations of
0\secpoint5--0\secpoint6 from Q1549-C25 in the $H_{160}$ image (one of which also corresponds
to a positive fluctuation in the $J_{125}$ image), these are not significant
and have no counterparts in $U_{336}$ or $V_{606}$.
\citet{mostardi2015} performed a detailed analysis of the multi-wavelength morphologies
and photometry of 16 LBGs in the HS1549+1919 field covered by 
$U_{336}V_{606}J_{125}H_{160}$ imaging, using the
$V_{606}$ image for detecting objects and defining isophotes with {\it SExtractor}
\citep{bertin1996}. In this analysis, Q1549-C25 was described by a single
segment, with uniform morphology in $V_{606}$, $J_{125}$, and $H_{160}$, and classified
as ``uncontaminated."

The $U_{336}$ filter provides a clean probe of the LyC spectral region at $z=3.15$,
and therefore a potential window on the morphology of escaping LyC emission.
However, as shown in Figure~\ref{fig:c25-panelfig}, Q1549-C25 is undetected in $U_{336}$.
This non-detection is entirely consistent with the spectroscopic detection of LyC
emission, given the depth of the $U_{336}$ image. Using an isophote defined by $V_{606}$,
we measure a 3$\sigma$ upper limit of $m_{336}=26.80$, consistent with the LRIS LyC detection,
based on the assumption of a flat spectrum between rest-frame $880-910$\AA\ and the effective rest
wavelength of the F336W filter, i.e., 808\AA\ (which is the most optimistic case, given
the likely increased IGM attenuation at shorter wavelengths). 

\subsection{Stellar Population Modeling}
\label{sec:results-sed}
\begin{figure}
\centering
\includegraphics[width=0.5\textwidth]{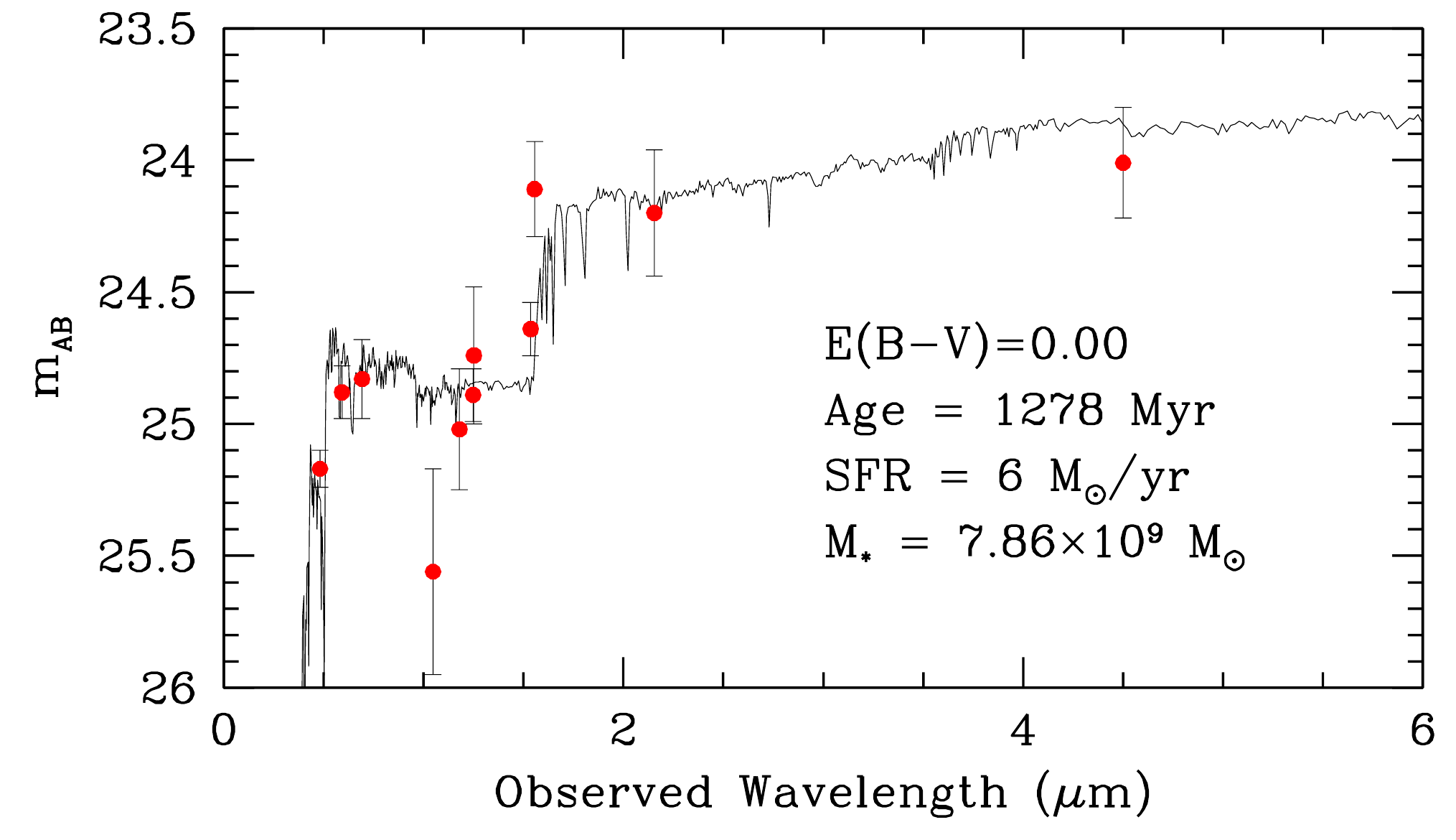}
\caption{Observed and best-fit model SEDs for Q1549-C25. $G$ and $V_{606}$
have been corrected for  Ly$\alpha$ emission, while $K_s$ has been corrected
for emission from [OIII]$\lambda 5007$ and H$\beta$.  The parameters
for the best-fit CSF model are indicated in the legend.
}
\label{fig:c25-sedplot}
\end{figure}

We used the \citet{bruzual2003} population synthesis code to
model the spectral energy distribution (SED) of Q1549-C25,
characterizing its stellar population and dust content.
For such modeling, we fit ground-based 
$G$, ${\cal R}$, $J_1$, $J_2$, $J_3$, $H_{short}$, along with
{\it HST} $V_{606}$, $J_{125}$, and $H_{160}$ and {\it Spitzer}/IRAC channel 2,
correcting $G$ and $V_{606}$ beforehand
for the contribution from Ly$\alpha$ emission ($W_{Ly\alpha,0}=15$\AA),
and $K_s$ for the combined contribution of [OIII] and H$\beta$ ($W_{(\mbox{[OIII]+H}\beta),0}=256$\AA).
Based on a constant star formation (CSF), solar-metallicity model, 
Chabrier IMF, and \citet{calzetti2000} extinction curve, 
we find $E(B-V)=0.0^{+0.0}_{-0.0}$,
Age$=1.3^{+0.5}_{-0.4}$~Gyr, SFR=$6^{+0}_{-1}\mbox{ }M_{\odot}\mbox{ yr}^{-1}$, and 
$M_*=7.9^{+2.5}_{-2.3}\times10^9 M_{\odot}$, where parameter confidence intervals reflect
the photometric uncertainties.\footnote{91\% of Monte Carlo realizations of the best-fit
model yielded $E(B-V)=0$.} Assuming an exponentially-rising star-formation
history yields very similar best-fit parameters. Figure~\ref{fig:c25-sedplot}
shows the SED of Q1549-C25, along with the best-fit CSF model.

\section{Discussion}
\label{sec:discussion}

\subsection{The LyC Escape Fraction}
\label{sec:discussion-fesc}

\begin{figure*}
\centering
\includegraphics[width=0.95\textwidth]{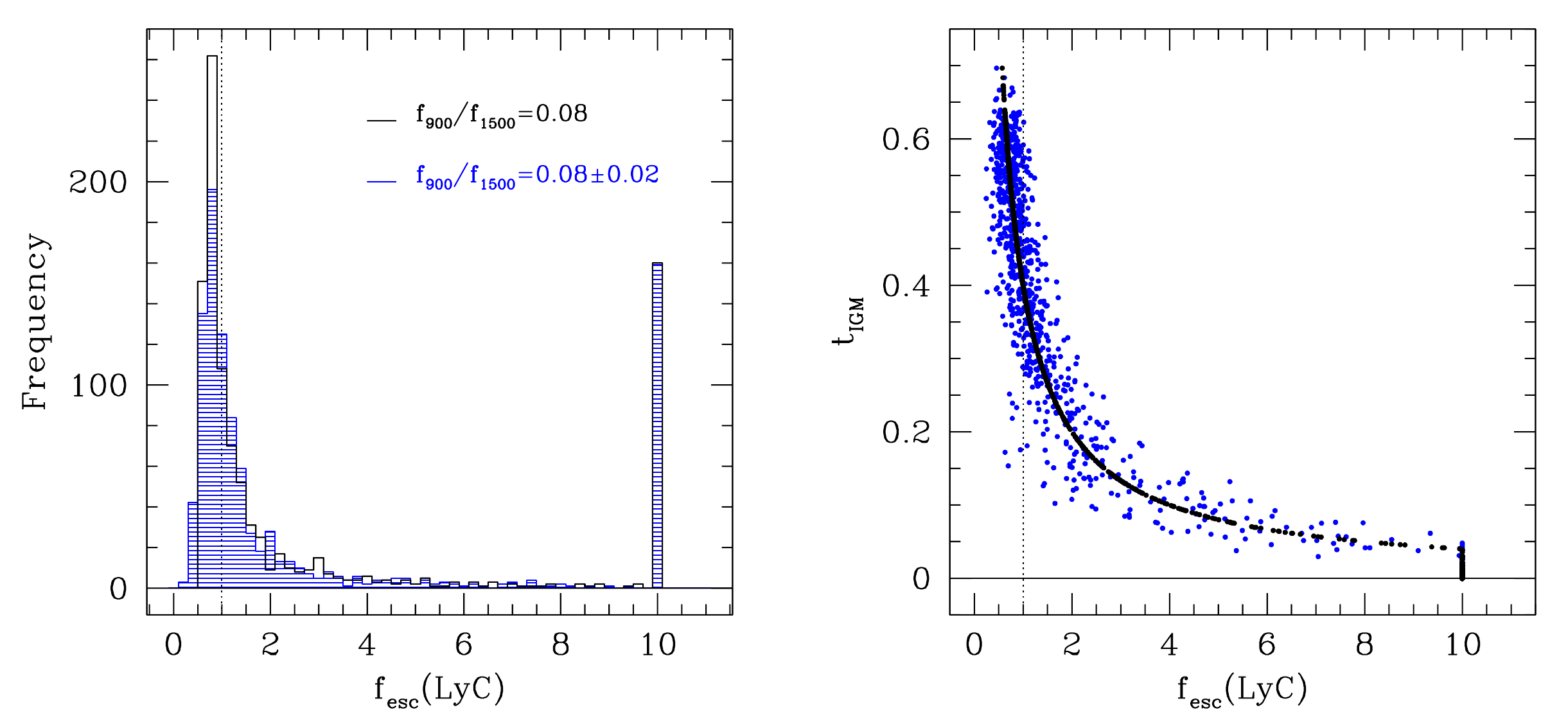}
\caption{ Distribution of $f_{\mbox{esc}}(\mbox{LyC})$ for Q1549-C25. 
$f_{\mbox{esc}}(\mbox{LyC})$ is estimated  based on the observed $f_{900}/f_{1500}$,
an assumed intrinsic ratio, $L_{900}/L_{1500}=0.20$, and 1000 simulated realizations
of IGM+CGM transmission in the LyC region at $z=3.15$. In each panel, black
symbols result from assuming no measurement uncertainty in $f_{900}/f_{1500}$, while
blue ones assume the observed error of 0.02. Values of $f_{\mbox{esc}}(\mbox{LyC}) > 10$
(i.e., resulting from low $t_{IGM}$) have been fixed at 10, and 
$f_{\mbox{esc}}(\mbox{LyC})=1$ is indicated as a vertical dotted line. 
{\bf Left:} Histogram of $f_{\mbox{esc}}(\mbox{LyC})$ for Q1549-C25. 
$f_{\mbox{esc}}(\mbox{LyC})\leq 1$ corresponds to 45\% of the realizations.
{\bf Right:} Joint distribution of $t_{IGM}$ vs. $f_{\mbox{esc}}(\mbox{LyC})$.
With the assumption of no error bar in $f_{900}/f_{1500}$, there is a one-to-one relation
between $t_{IGM}$ and $f_{\mbox{esc}}(\mbox{LyC})$.
}
\label{fig:fescdistfigs}
\end{figure*}

We can estimate the LyC escape fraction of Q1549-C25 based
on the observed LyC to non-ionizing UV flux-density ratio, $f_{900}/f_{1500}$,
the intrinsic luminosity-density ratio, $L_{900}/L_{1500}$, and the IGM
transmission factor, $t_{IGM}$. The escape fraction is typically quoted
in both {\it relative} and {\it absolute} terms. The relative escape
fraction, $f_{\mbox{esc,rel}}(\mbox{LyC})$ is a measure of how the observed
$f_{900}/f_{1500}$ ratio (corrected for IGM absorption) compares to the 
theoretical one, while the absolute escape fraction, $f_{\mbox{esc}}(\mbox{LyC})$,
is simply the ratio of the escaping to intrinsic LyC luminosity density. In terms
of the quantities described above, we define $f_{\mbox{esc,rel}}(\mbox{LyC})$
as:

\begin{equation}
f_{\mbox{esc,rel}}(\mbox{LyC})=\frac{(f_{900}/f_{1500})}{(L_{900}/L_{1500})t_{IGM}}
\label{eq:fescrel}
\end{equation}

The absolute escape fraction is defined as:

\begin{equation}
f_{\mbox{esc}}(\mbox{LyC})=f_{\mbox{esc,rel}}(\mbox{LyC})\times f_{\mbox{esc}}(\mbox{1500})
\label{eq:fescdef}
\end{equation}

where $f_{\mbox{esc}}(\mbox{1500})$ refers to the absolute escape fraction at $1500$\AA,
modulated by dust attenuation. Since $E(B-V)=0$ for Q1549-C25, $f_{\mbox{esc}}(\mbox{1500})=1$,
and $f_{\mbox{esc}}(\mbox{LyC}) = f_{\mbox{esc,rel}}(\mbox{LyC})$. Therefore,
equation~\ref{eq:fescrel} can be rewritten as: 

\begin{equation}
f_{900}/f_{1500} = f_{\mbox{esc}}(\mbox{LyC})(L_{900}/L_{1500})t_{IGM}
\label{eq:f900f1500}
\end{equation}

Our measurement of $f_{900}/f_{1500}=0.08\pm 0.02$ constrains
the quantity, $f_{\mbox{esc}}(\mbox{LyC})(L_{900}/L_{1500})t_{IGM}$.
A precise estimate of $f_{\mbox{esc}}(\mbox{LyC})$ is not possible, given
the uncertainties in $t_{IGM}$ for a single $z=3.15$ sightline \citep{vanzella2016},
and, to a lesser extent, $L_{900}/L_{1500}$. However, based on reasonable
assumptions, we can place rough constraints on $f_{\mbox{esc}}(\mbox{LyC})$.

In order to investigate constraints on $f_{\mbox{esc}}(\mbox{LyC})$, we generated 1000
model $z=3.15$ sightlines simulating the distribution of H~I absorbers in both the IGM
and circumgalactic medium (CGM), the latter reflecting the the enhancement in intergalactic H~I absorption
in the vicinity of high-redshift star-forming galaxies.
The absorber distributions were based on observations from
\citet{rudie2012} and \citet{rudie2013}, and are described in more detail in
Steidel et al. (in prep). For each model sightline, we calculated the average
transmission factor, $t_{IGM}$, over the rest wavelength range, $880-910$\AA.
We also considered the intrinsic luminosity-density ratio, $L_{900}/L_{1500}$,
with $L_{900}$ and $L_{1500}$ defined in units of $\mbox{ergs s}^{-1}\mbox {Hz}^{-1}$. 
For an age of $\sim 1$~Gyr, solar-metallicity
CSF models predict $L_{900}/L_{1500}$ ranging from $0.14-0.25$
\citep{bruzual2003,siana2007,stanway2016}.
Larger values are possible for lower metallicities
or different assumptions regarding massive stellar evolution. We adopted $L_{900}/L_{1500}=0.20$ as a fiducial value.

Based on the ensemble of simulated $z=3.15$ sightlines, the measurement and uncertainty
in $f_{900}/f_{1500}$, and the assumed value of $L_{900}/L_{1500}=0.20$, we calculated
the distribution of $f_{\mbox{esc}}(\mbox{LyC})$ (Figure~\ref{fig:fescdistfigs}, left,
blue shaded histogram). We find that slightly less than half (45\%) of the 
distribution of $f_{\mbox{esc}}(\mbox{LyC})$
is at $\leq 1$, and 95\% of the distribution is at $f_{\mbox{esc}}(\mbox{LyC}) \geq 0.51$.
The results are very similar if, instead of assuming $L_{900}/L_{1500}=0.20$,
we randomly draw $L_{900}/L_{1500}$ values uniformly between 0.14 and 0.25.
It is also useful to visualize the inverse relationship between $t_{IGM}$ and $f_{\mbox{esc}}(\mbox{LyC})$,
given $f_{900}/f_{1500}$ and $L_{900}/L_{1500}$ (Figure~\ref{fig:fescdistfigs}, right).
With no measurement error (black points), a requirement of $f_{\mbox{esc}}(\mbox{LyC}) \leq 1$
translates into an IGM transmission factor of $t_{IGM} \geq 0.40$, which is higher
than the average at $z=3.15$. Scatter is introduced into the $t_{IGM}$ vs. $f_{\mbox{esc}}(\mbox{LyC})$
relationship by the measurement uncertainty in $f_{900}/f_{1500}$ (blue points), yet the requirement
of $f_{\mbox{esc}}(\mbox{LyC}) \leq 1$ implies a mean transmission of $\langle t_{IGM} \rangle=0.519$,
which is significantly higher
than through a random $z=3.15$ sightline ($\langle t_{IGM}(z=3.15) \rangle = 0.339$).

\subsection{A Comparison of Q1549-C25 and {\it Ion2}}
\label{sec:discussion-c25comp}

The only other $z\sim 3$ galaxy with a direct spectroscopic detection of uncontaminated
LyC emission is {\it Ion2} at $z=3.21$ \citep{debarros2016,vanzella2016}. It is therefore
useful to compare the properties of {\it Ion2} and Q1549-C25.
In terms of the escape of LyC emission, {\it Ion2} appears more extreme than Q1549-C25.
\citet{debarros2016} report $f_{900}/f_{1500}=0.11$ ($\mbox{S/N}\sim 5$)
for {\it Ion2}.
Using the same IGM+CGM model tuned to $z=3.2$, and the same assumed value for $L_{900}/L_{1500}$,
we find that 95\% of the distribution of $f_{\mbox{esc}}(\mbox{LyC})$ is at $\geq 0.79$ for
{\it Ion2}. {\it Ion2}, like Q1549-C25, is characterized by negligible dust extinction,
and therefore  $f_{\mbox{esc,rel}}(\mbox{LyC})= f_{\mbox{esc}}(\mbox{LyC})$.
\citet{debarros2016} report a significantly lower escape fraction for {\it Ion2}
($f_{\mbox{esc}}(\mbox{LyC})=0.64\pm 0.1$), but assume
a more transparent model for IGM opacity \citep{inoue2014}, and $L_{900}/L_{1500}=1/3$.
Regardless of the particular IGM model assumed, both sources are characterized
by significantly higher $f_{900}/f_{1500}$ values and $f_{\mbox{esc}}(\mbox{LyC})$  distributions
than average at $z\sim 3$ (Steidel et al., in prep), suggesting large variations in these quantities
among high-redshift star-forming galaxies.

Along with a higher $f_{900}/f_{1500}$, {\it Ion2} also has significantly larger
rest-frame Ly$\alpha$ and [OIII]$\lambda 5007$ emission equivalent widths than Q1549-C25, 
with  $W_{Ly\alpha,0}=94$~\AA\ and $W_{\mbox{[OIII]},0}=1500$~\AA.  
In addition, no low-ionization metal absorption lines are detected in the rest-UV spectrum
of {\it Ion2}, suggesting a lower covering fraction of neutral gas than the spectrum
of Q1549-C25, in which several low-ionization lines are detected. Finally,
the Ly$\alpha$ profile of {\it Ion2} is double-peaked, with the centroid of the blue peak
coinciding with the galaxy systemic redshift. Comparing this profile with Ly$\alpha$
radiative transfer models, \citet{debarros2016} cite it as evidence for a low neutral
gas column density. The Ly$\alpha$ profile of Q1549-C25 on the other hand is described by a single
peak (at a resolution of $R\sim 1400$, higher than that of the VLT/VIMOS
spectrum of {\it Ion2}), which is redshifted by $500\mbox{ km s}^{-1}$
relative to the interstellar absorption lines, and $250\mbox{ km s}^{-1}$
with respect to the systemic redshift. The large-scale outflow in Q1549-C25, reflected
by the offsets of Ly$\alpha$ and interstellar
absorption redshifts \citep{shapley2003} relative to systemic, may yield a porous ISM whose gaps
provide escape routes for LyC photons.

As stated above, the stellar population fits for both galaxies imply $E(B-V)\sim0$,
which is conducive to the escape of LyC radiation, and SFRs and stellar masses
lower than the median for ${\cal R}\leq25.5$ LBGs ($\mbox{SFR }=6M_{\odot}\mbox{ yr}^{-1}$ 
and $M_*=7.9\times10^{9}M_{\odot}$ for Q1549-C25, and $\mbox{SFR }=16M_{\odot}\mbox{ yr}^{-1}$
and $M_*=3.2\times10^{9}M_{\odot}$ for {\it Ion2}).
The best-fit age for Q1549-C25 is $\sim 1$~Gyr,
while it is 400~Myr for {\it Ion2}. Both of these are older than the median CSF age
derived for LBGs \citep{kornei2010}, which is notable, given that the intrinsic ratio of
LyC to non-ionizing UV luminosity density, $L_{900}/L_{1500}$, is highest
at young ages ($\leq 10$~Myr) and declines to a minimum value at ages $>300$~Myr
\citep[assuming continuous star formation;][]{siana2007}. \citet{mostardi2015}
present photometric evidence that the galaxy, Q1549-MD5b is leaking LyC radiation, and 
described by an age of only 50~Myr. Accordingly, LyC leakage appears to occur
over a wide range in galaxy age.

\subsection{Outlook}
\label{sec:discussion-outlook}
The direct and uncontaminated spectroscopic detection of LyC emission has
now been achieved for two galaxies at $z\sim 3$. Based on reasonable assumptions,
both sources suggest high escape fractions ($\gtrsim0.5$), though the constraints
on $f_{\mbox{esc}}(\mbox{LyC})$ are not precise. Both sources are characterized
by negligible dust extinction and strong Ly$\alpha$ emission, and, contrary to
simple expectations, stellar population ages older than the median for $z\sim3$ LBGs.
In order to make progress on estimating the typical LyC escape fraction at $z\sim 3$ and during the epoch
of reionization, we now require an order-of-magnitude larger sample of galaxies with 
clean detections of LyC emission. With such a sample, averaged over many sightlines,
the constraints on the mean IGM transmission and therefore $f_{\mbox{esc}}(\mbox{LyC})$
will be much stronger. Our survey
of LBGs with deep LRIS spectroscopy of the LyC region has yielded 13 galaxies,
including Q1549-C25, with apparent spectroscopic detections of LyC and no evidence of blending
from lower-redshift interlopers.
Observations at the spatial resolution of {\it HST} will enable us to rule out possible contamination
and characterize the global contribution of star-forming galaxies at $z\geq 3$
to the ionizing background.

\section*{Acknowledgements}
We thank the anonymous referee for a constructive report.
CCS acknowledges support from NSF grants AST-0908805 and AST-1313472.
AES acknowledges support from the David \& Lucile Packard Foundation.
NAR is supported by an Alfred P. Sloan Research Fellowship. 
We wish to extend special thanks to those of Hawaiian ancestry on
whose sacred mountain we are privileged to be guests. Without their generous hospitality, most
of the observations presented herein would not have been possible.


\begin{thebibliography}{}
\expandafter\ifx\csname natexlab\endcsname\relax\def\natexlab#1{#1}\fi

\bibitem[{{Bertin} \& {Arnouts}(1996)}]{bertin1996}
{Bertin}, E., \& {Arnouts}, S. 1996, \aaps, 117, 393

\bibitem[{{Borthakur} {et~al.}(2014){Borthakur}, {Heckman}, {Leitherer}, \&
  {Overzier}}]{borthakur2014}
{Borthakur}, S., {Heckman}, T.~M., {Leitherer}, C., \& {Overzier}, R.~A. 2014,
  Science, 346, 216

\bibitem[{{Bruzual} \& {Charlot}(2003)}]{bruzual2003}
{Bruzual}, G., \& {Charlot}, S. 2003, \mnras, 344, 1000

\bibitem[{{Calzetti} {et~al.}(2000){Calzetti}, {Armus}, {Bohlin}, {Kinney},
  {Koornneef}, \& {Storchi-Bergmann}}]{calzetti2000}
{Calzetti}, D., {Armus}, L., {Bohlin}, R.~C., {et~al.} 2000, \apj, 533, 682

\bibitem[{{de Barros} {et~al.}(2016){de Barros}, {Vanzella}, {Amor{\'{\i}}n},
  {Castellano}, {Siana}, {Grazian}, {Suh}, {Balestra}, {Vignali}, {Verhamme},
  {Zamorani}, {Mignoli}, {Hasinger}, {Comastri}, {Pentericci},
  {P{\'e}rez-Montero}, {Fontana}, {Giavalisco}, \& {Gilli}}]{debarros2016}
{de Barros}, S., {Vanzella}, E., {Amor{\'{\i}}n}, R., {et~al.} 2016, \aap, 585,
  A51

\bibitem[{{Finkelstein} {et~al.}(2012){Finkelstein}, {Papovich}, {Ryan},
  {Pawlik}, {Dickinson}, {Ferguson}, {Finlator}, {Koekemoer}, {Giavalisco},
  {Cooray}, {Dunlop}, {Faber}, {Grogin}, {Kocevski}, \&
  {Newman}}]{finkelstein2012}
{Finkelstein}, S.~L., {Papovich}, C., {Ryan}, R.~E., {et~al.} 2012, \apj, 758,
  93

\bibitem[{{Grazian} {et~al.}(2016){Grazian}, {Giallongo}, {Gerbasi}, {Fiore},
  {Fontana}, {Le F{\`e}vre}, {Pentericci}, {Vanzella}, {Zamorani}, {Cassata},
  {Garilli}, {Le Brun}, {Maccagni}, {Tasca}, {Thomas}, {Zucca},
  {Amor{\'{\i}}n}, {Bardelli}, {Cassar{\`a}}, {Castellano}, {Cimatti},
  {Cucciati}, {Durkalec}, {Giavalisco}, {Hathi}, {Ilbert}, {Lemaux}, {Paltani},
  {Ribeiro}, {Schaerer}, {Scodeggio}, {Sommariva}, {Talia}, {Tresse},
  {Vergani}, {Bonchi}, {Boutsia}, {Capak}, {Charlot}, {Contini}, {de la Torre},
  {Dunlop}, {Fotopoulou}, {Guaita}, {Koekemoer}, {L{\'o}pez-Sanjuan},
  {Mellier}, {Merlin}, {Paris}, {Pforr}, {Pilo}, {Santini}, {Scoville},
  {Taniguchi}, \& {Wang}}]{grazian2016}
{Grazian}, A., {Giallongo}, E., {Gerbasi}, R., {et~al.} 2016, \aap, 585, A48

\bibitem[{{Inoue} {et~al.}(2014){Inoue}, {Shimizu}, {Iwata}, \&
  {Tanaka}}]{inoue2014}
{Inoue}, A.~K., {Shimizu}, I., {Iwata}, I., \& {Tanaka}, M. 2014, \mnras, 442,
  1805

\bibitem[{{Izotov} {et~al.}(2016){Izotov}, {Orlitov{\'a}}, {Schaerer}, {Thuan},
  {Verhamme}, {Guseva}, \& {Worseck}}]{izotov2016}
{Izotov}, Y.~I., {Orlitov{\'a}}, I., {Schaerer}, D., {et~al.} 2016, \nat, 529,
  178

\bibitem[{{Kornei} {et~al.}(2010){Kornei}, {Shapley}, {Erb}, {Steidel},
  {Reddy}, {Pettini}, \& {Bogosavljevi{\'c}}}]{kornei2010}
{Kornei}, K.~A., {Shapley}, A.~E., {Erb}, D.~K., {et~al.} 2010, \apj, 711, 693

\bibitem[{{Kuhlen} \& {Faucher-Gigu{\`e}re}(2012)}]{kuhlen2012}
{Kuhlen}, M., \& {Faucher-Gigu{\`e}re}, C.-A. 2012, \mnras, 423, 862

\bibitem[{{Leitherer} {et~al.}(2016){Leitherer}, {Hernandez}, {Lee}, \&
  {Oey}}]{leitherer2016}
{Leitherer}, C., {Hernandez}, S., {Lee}, J.~C., \& {Oey}, M.~S. 2016, ArXiv
  e-prints, arXiv:1603.06779

\bibitem[{{Mostardi} {et~al.}(2013){Mostardi}, {Shapley}, {Nestor}, {Steidel},
  {Reddy}, \& {Trainor}}]{mostardi2013}
{Mostardi}, R.~E., {Shapley}, A.~E., {Nestor}, D.~B., {et~al.} 2013, \apj, 779,
  65

\bibitem[{{Mostardi} {et~al.}(2015){Mostardi}, {Shapley}, {Steidel}, {Trainor},
  {Reddy}, \& {Siana}}]{mostardi2015}
{Mostardi}, R.~E., {Shapley}, A.~E., {Steidel}, C.~C., {et~al.} 2015, \apj,
  810, 107

\bibitem[{{Nestor} {et~al.}(2011){Nestor}, {Shapley}, {Steidel}, \&
  {Siana}}]{nestor2011}
{Nestor}, D.~B., {Shapley}, A.~E., {Steidel}, C.~C., \& {Siana}, B. 2011, \apj,
  736, 18

\bibitem[{{Pettini} {et~al.}(2001){Pettini}, {Shapley}, {Steidel}, {Cuby},
  {Dickinson}, {Moorwood}, {Adelberger}, \& {Giavalisco}}]{pettini2001}
{Pettini}, M., {Shapley}, A.~E., {Steidel}, C.~C., {et~al.} 2001, \apj, 554,
  981

\bibitem[{{Reddy} {et~al.}(2012){Reddy}, {Pettini}, {Steidel}, {Shapley},
  {Erb}, \& {Law}}]{reddy2012}
{Reddy}, N.~A., {Pettini}, M., {Steidel}, C.~C., {et~al.} 2012, \apj, 754, 25

\bibitem[{{Reddy} {et~al.}(2008){Reddy}, {Steidel}, {Pettini}, {Adelberger},
  {Shapley}, {Erb}, \& {Dickinson}}]{reddy2008}
{Reddy}, N.~A., {Steidel}, C.~C., {Pettini}, M., {et~al.} 2008, \apjs, 175, 48

\bibitem[{{Robertson} {et~al.}(2015){Robertson}, {Ellis}, {Furlanetto}, \&
  {Dunlop}}]{robertson2015}
{Robertson}, B.~E., {Ellis}, R.~S., {Furlanetto}, S.~R., \& {Dunlop}, J.~S.
  2015, \apjl, 802, L19

\bibitem[{{Rudie} {et~al.}(2013){Rudie}, {Steidel}, {Shapley}, \&
  {Pettini}}]{rudie2013}
{Rudie}, G.~C., {Steidel}, C.~C., {Shapley}, A.~E., \& {Pettini}, M. 2013,
  \apj, 769, 146

\bibitem[{{Rudie} {et~al.}(2012){Rudie}, {Steidel}, {Trainor}, {Rakic},
  {Bogosavljevi{\'c}}, {Pettini}, {Reddy}, {Shapley}, {Erb}, \&
  {Law}}]{rudie2012}
{Rudie}, G.~C., {Steidel}, C.~C., {Trainor}, R.~F., {et~al.} 2012, \apj, 750,
  67

\bibitem[{{Schlegel} {et~al.}(1998){Schlegel}, {Finkbeiner}, \&
  {Davis}}]{schlegel1998}
{Schlegel}, D.~J., {Finkbeiner}, D.~P., \& {Davis}, M. 1998, \apj, 500, 525

\bibitem[{{Shapley} {et~al.}(2003){Shapley}, {Steidel}, {Pettini}, \&
  {Adelberger}}]{shapley2003}
{Shapley}, A.~E., {Steidel}, C.~C., {Pettini}, M., \& {Adelberger}, K.~L. 2003,
  \apj, 588, 65

\bibitem[{{Shapley} {et~al.}(2006){Shapley}, {Steidel}, {Pettini},
  {Adelberger}, \& {Erb}}]{shapley2006}
{Shapley}, A.~E., {Steidel}, C.~C., {Pettini}, M., {Adelberger}, K.~L., \&
  {Erb}, D.~K. 2006, \apj, 651, 688

\bibitem[{{Siana} {et~al.}(2007){Siana}, {Teplitz}, {Colbert}, {Ferguson},
  {Dickinson}, {Brown}, {Conselice}, {de Mello}, {Gardner}, {Giavalisco}, \&
  {Menanteau}}]{siana2007}
{Siana}, B., {Teplitz}, H.~I., {Colbert}, J., {et~al.} 2007, \apj, 668, 62

\bibitem[{{Siana} {et~al.}(2015){Siana}, {Shapley}, {Kulas}, {Nestor},
  {Steidel}, {Teplitz}, {Alavi}, {Brown}, {Conselice}, {Ferguson}, {Dickinson},
  {Giavalisco}, {Colbert}, {Bridge}, {Gardner}, \& {de Mello}}]{siana2015}
{Siana}, B., {Shapley}, A.~E., {Kulas}, K.~R., {et~al.} 2015, \apj, 804, 17

\bibitem[{{Stanway} {et~al.}(2016){Stanway}, {Eldridge}, \&
  {Becker}}]{stanway2016}
{Stanway}, E.~R., {Eldridge}, J.~J., \& {Becker}, G.~D. 2016, \mnras, 456, 485

\bibitem[{{Steidel} {et~al.}(2003){Steidel}, {Adelberger}, {Shapley},
  {Pettini}, {Dickinson}, \& {Giavalisco}}]{steidel2003}
{Steidel}, C.~C., {Adelberger}, K.~L., {Shapley}, A.~E., {et~al.} 2003, \apj,
  592, 728

\bibitem[{{Steidel} {et~al.}(2014){Steidel}, {Rudie}, {Strom}, {Pettini},
  {Reddy}, {Shapley}, {Trainor}, {Erb}, {Turner}, {Konidaris}, {Kulas}, {Mace},
  {Matthews}, \& {McLean}}]{steidel2014}
{Steidel}, C.~C., {Rudie}, G.~C., {Strom}, A.~L., {et~al.} 2014, \apj, 795, 165

\bibitem[{{Vanzella} {et~al.}(2012){Vanzella}, {Guo}, {Giavalisco}, {Grazian},
  {Castellano}, {Cristiani}, {Dickinson}, {Fontana}, {Nonino}, {Giallongo},
  {Pentericci}, {Galametz}, {Faber}, {Ferguson}, {Grogin}, {Koekemoer},
  {Newman}, \& {Siana}}]{vanzella2012}
{Vanzella}, E., {Guo}, Y., {Giavalisco}, M., {et~al.} 2012, \apj, 751, 70

\bibitem[{{Vanzella} {et~al.}(2016){Vanzella}, {de Barros}, {Vasei}, {Alavi},
  {Giavalisco}, {Siana}, {Grazian}, {Hasinger}, {Suh}, {Cappelluti}, {Vito},
  {Amorin}, {Balestra}, {Brusa}, {Calura}, {Castellano}, {Comastri}, {Fontana},
  {Gilli}, {Mignoli}, {Pentericci}, {Vignali}, \& {Zamorani}}]{vanzella2016}
{Vanzella}, E., {de Barros}, S., {Vasei}, K., {et~al.} 2016, ArXiv e-prints,
  arXiv:1602.00688

\end{thebibliography}

\end{document}